\documentclass[11pt,aps,nofootinbib,nolongbibliography]{revtex4-2}
\pdfoutput=1

\usepackage{setspace}
\usepackage{url}
\usepackage{hyperref}
\hypersetup{
	colorlinks=true,
	urlcolor=blue,
	linkcolor=blue,
	citecolor=blue
}

\usepackage{amsmath}
\usepackage{amssymb}
\usepackage{bbm}
\usepackage{mathtools}
\usepackage[normalem]{ulem}
\usepackage{dsfont}
\usepackage{mathrsfs}
\usepackage[makeroom]{cancel}
\usepackage{graphicx}
\usepackage{bm}

\usepackage[T1]{fontenc}
\usepackage[ddmmyyyy]{datetime}
\makeatletter
\def\Dated@name{}
\makeatother

\def\B#1{\left(#1\right)}

\def\FOR{{ \ \text{for} \ } }
\def\AND{{ \ \text{and} \ } }
\DeclareMathOperator\sgn{\text{sgn}}

\newcommand{\Rzym}[1]{%
  \textup{\uppercase\expandafter{\romannumeral#1}}%
}

\begin{document}

\title{Lorentz transformations in $1+1$ dimensional spacetime: mainly the
superluminal case}
\author{Bogdan S. Damski}
\affiliation{Jagiellonian University, 
Faculty of Physics, Astronomy and Applied Computer Science,
{\L}ojasiewicza 11, 30-348 Krak\'ow, Poland}
\begin{abstract}
We discuss the most general form of the Lorentz transformation
in $1+1$ dimensional spacetime, focusing mainly on its
superluminal branch. For this purpose, we introduce the 
 $2$-velocity of a reference frame and the clockwork postulate.
Basic special relativity effects are discussed in the proposed
framework. Different forms of the superluminal Lorentz
transformation, which were studied  
 in the literature, are critically  examined from the perspective
 of our formalism. Counterintuitive features of the superluminal
 Lorentz transformation are identified  both in our 
 approach and in earlier studies.
\end{abstract}
\maketitle

\section{Introduction}

The topic of superluminal particles, which are nowadays called tachyons (term coined by Feinberg \cite{feinberg}), is still controversial among physicists for two reasons. The first one is that the existence of tachyons carrying information is commonly believed to lead to violations of causality \cite{tolman}. The second one is that no experimental evidence has been found for their existence. On the theoretical side,  different attempts of constructing a legitimate theory were made. However, none of them was accepted by the whole community. On the experimental side, some physicists claimed that they registered tachyons, but the result of their experiment was either not repeatable or it was later shown that a mistake was made in the measurement process (see \cite{ehrlich} for review of such efforts).

The recent publication of Dragan and Ekert has once again renewed interest in tachyon-related physics \cite{dragan}.  In their work, they proposed that the theory of relativity and quantum mechanics are deeply connected. They argued that the violations of causality associated with  tachyons  explain the probabilistic nature of quantum effects. According to them, the presence of tachyons in the theory does not lead to causal paradoxes as in \cite{tolman}, but rather changes our notion  of causality. Their novel ideas triggered a heated debate in the scientific community \cite{grudka_2022,grudka_2023,Horodecki_2023,DelSanto_2022, Dragan_2022_reply_Grudka,Dragan_2023_reply_Horodecki,Dragan_2023_reply_Horvat}.

In traditional relativity, where only subluminal reference frames are considered, a tachyon is a particle that is never at rest. Thereby, various authors considered the concept of superluminal reference frames. Such studies date back at least to the work of Parker \cite{parker}, where 
the Lorentz transformation has the following form 
\begin{subequations}
\begin{align}
&|u|<1: \  t' = \gamma(u)(t-u x), \quad x'=\gamma(u)(x-u t),\\
&|u|>1:  \ t' = -\sgn(u)\gamma(u)(t-u x), \quad x'=-\sgn(u)\gamma(u)(x-ut)
\label{nad}
\end{align}
\label{superIIIgen}%
\end{subequations}
and the $-\sgn(u)\to\sgn(u)$ version of (\ref{superIIIgen}) is said to lead to the same physics ($\sgn(x)=0,\pm1$ is 
the sign function).
According to these equations, the primed reference
frame moves with the dimensionless velocity $u$ relative to the unprimed one (the velocity is measured in the units of the speed of light) and $\gamma(u)=1/|1-u^2|^{1/2}$. We say that the  velocity $u$ 
is subluminal for $|u|<1$ and superluminal for $|u|>1$.

Besides (\ref{superIIIgen}), there are other versions  of 
the Lorentz transformation present in the literature such as 
\begin{equation}
u\in\mathbbm{R}\setminus\{1,-1\}: \  t' = \gamma(u)(t-u x), \quad x'=\gamma(u)(x-u t)
\label{Itr}
\end{equation}
or its $\gamma(u)\to\sgn(1-u^2)\gamma(u)$ version  (see e.g. 
\cite{goldoni}, references cited in \cite{Antippa1983}, and \cite{henriksen}).
However, 
it can be shown that none of them is actually acceptable \cite{Antippa1983}, which 
illustrates the fact that the extension of the standard 
Lorentz transformation to the superluminal regime 
involves some subtle issues.

In this work, we will introduce the  $2$-velocity of
a reference frame and the clockwork postulate in Sec. \ref{2_sec}, 
which will allow us
to present  superluminal Lorentz 
transformations from a different angle. 
We will examine in Secs. \ref{3_sec} and \ref{length_sec}
the new formalism 
 in the context of  basic special  relativity 
effects. 
The discussion of (\ref{superIIIgen}) and (\ref{Itr}) from the perspective
of our formalism 
will be presented in Sec. 
\ref{5_sec}. The summary of our work will be 
given in Sec. \ref{6_sec}.

\section{$2$-velocity of reference frame}
\label{2_sec}
When one thinks  about the Lorentz transformation in $1+1$ 
dimensional spacetime, one 
considers two reference frames moving relative to each other 
with some  velocity, say $u$. 
Then, one uses  such a velocity to 
parametrize the coefficients in the transformation
relating the  spacetime coordinates 
in the two reference frames.
Such coefficients take the form 
\begin{equation}
\pm \gamma(u), \ \pm \gamma(u)|u|,
\label{ab}
\end{equation}
where the signs in both expressions are independently chosen.
While there is in principle nothing wrong with such a procedure,
we find it hardly satisfactory for the following reason. Namely, 
the special theory of relativity is about  physics happening 
in  spacetime, where the time-like and space-like 
features appear on {\it equal footing} in different contexts.
Thereby, it is our opinion that  it would be
more natural to use the $2$-velocity to characterize 
the relation between the spacetime coordinates in the 
two reference frames.
Such an observation  also naturally follows from (\ref{ab}),
which   suggests the consideration of the $2$-vector 
\begin{equation}
U=(\pm \gamma(u),\pm  \gamma(u)|u|),
\label{Upm}
\end{equation}
which is reminiscent of the  $2$-velocity of a relativistic particle.
The basic property of (\ref{Upm}) is  
that 
\begin{equation}
U\cdot U \equiv \B{U^0}^2 - \B{U^1}^2= \sgn(1-u^2)=
\left\{
\begin{array}{l}
+1  \ \FOR u  \ \text{subluminal}\\
-1   \ \FOR u  \ \text{superluminal}
\end{array}
\right.
\label{UUs}
\end{equation}
regardless of the choice of signs in (\ref{Upm}) (all dot products  are defined in this work
with the metric tensor $\text{diag}(1,-1)$; the signs in (\ref{Upm}) are independently 
chosen so that there are four expressions encoded in such an equation).

We propose the following parametrization of the  Lorentz transformation
\begin{equation}
t' = U\cdot U(U^0 t- U^1 x), \quad x'=U\cdot U(U^0 x- U^1 t),
\label{tprimuu}
\end{equation}
which leads to the following inverse transformation after the employment 
of (\ref{UUs})
\begin{equation}
t = U^0 t'+ U^1 x', \quad x=U^0 x'+ U^1 t'.
\label{tuu}
\end{equation}

\begin{figure}[t]
\includegraphics[width=0.75\columnwidth]{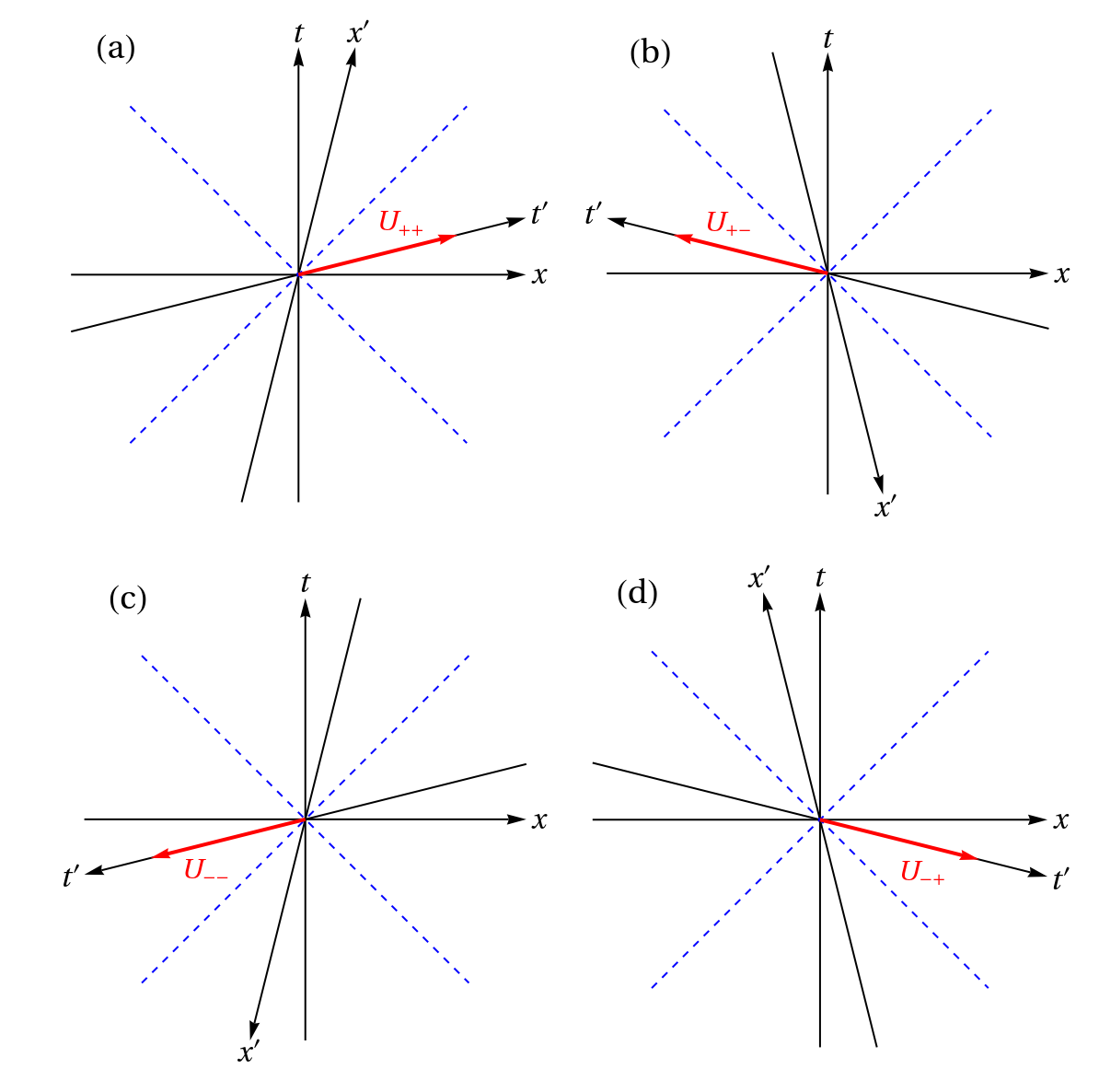}
\caption{The schematic illustration of four possibilities for 
orientation of the axes of the superluminal reference frames.
The red arrows display $2$-velocities 
$U_{\alpha\beta}=(\alpha\gamma(u),\beta\gamma(u)|u|)$, where 
$\alpha,\beta=\pm$ and $|u|>1$.
The light cones are plotted with blue dashed lines.}
\label{4osie}
\end{figure}

To give physical meaning to $U$, we remark that the 
orientation of the $t'$ axis on the spacetime diagram $(t,x)$
is given by $U$. Such an observation 
follows from the 
$x'=0$ version of (\ref{tuu}). The four superluminal 
options  resulting from 
different sign choices in (\ref{Upm}) are illustrated in  Fig. \ref{4osie}. 

{\bf Clockwork postulate.} In order to add meaning to the four possible orientations of the $t'$ axis, we introduce the \textit{clockwork postulate} according to which \textit{the proper time of an inertial observer always increases}. In other words, an observer always becomes older (never younger) in his own rest frame,
which nicely fits the basic assumption of  special relativity that all inertial observers and/or reference frames are equivalent.
This postulate implies that  the stationary observer in the primed reference frame always moves in the positive direction of the $t'$ axis. 
 The above-stated clockwork postulate 
should not be confused with the  clock postulate (also known as 
the clock hypothesis), which is based on the assumption
that the rate of operation of the ideal
clock in motion depends only on its instantaneous velocity (see \cite{DraganIdealClock} for 
a recent critical discussion of the clock postulate).

The remark formulated below (\ref{tuu}), along with the clockwork postulate, 
leads to the conclusion that $U$ is the
relativistic $2$-velocity of the primed
reference frame 
relative to the unprimed one.
Thereby, the  four options displayed in 
Fig. \ref{4osie} come from the fact that the
primed reference frame can be moving 
either forwards ($U^0>0$) or backwards 
($U^0<0$) in time $t$
and either in the positive ($U^1>0$) or in the 
negative ($U^1<0$) direction of the  $x$ axis.
In other words, all spacetime  
 options 
for the motion of the reference frame are 
encoded in $U$. Finally, we note that 
reference frames moving either 
forwards or backwards 
in time and in different directions in space  were
considered  in  $1+1$ ($1+3$) dimensional spacetime in 
the paper of Viera \cite{Viera} (Sutherland and Shepanski \cite{sutherland}).
However, the above introduced $U$-parametrization of these
transformations was  not explored in   \cite{Viera,sutherland}.

\section{Properties of $U$-parametrized transformation}
\label{3_sec}

We begin the discussion here from the derivation of the addition law for 
$2$-velocities.   We consider three inertial 
reference frames $O$, $O'$ and $O''$. 
We assume that $O'$ is moving with the $2$-velocity $V$ relative to $O$, 
$O''$ is moving with the $2$-velocity $U$ relative to $O$ 
and $U'$ relative to $O'$.
To determine $U'$, we note that 
such a $2$-velocity is defined via 
\begin{equation}
\binom{t'}{x'}=
\binom{U'^0 \quad U'^1}{U'^1 \quad U'^0}
\binom{t''}{x''},
\end{equation}
which can be compared to 
\begin{equation}
\binom{t'}{x'}=
V\cdot V\binom{V^0U^0-V^1U^1  \quad V^0U^1-V^1U^0}{V^0U^1-V^1U^0 \quad V^0U^0-V^1U^1}
\binom{t''}{x''}
\end{equation}
that is obtained via the transformation $O'\to O$ followed by the
$O\to O''$ transformation.\footnote{By the transformation
$O'\to O$ we understand here   transformation 
(\ref{tprimuu}) subjected to the replacement $U\to V$.}
This leads to the identification
\begin{equation}
\label{Uti}
U'=V\cdot V(V\cdot U, \varepsilon_{\mu\nu} V^\mu U^\nu), \ \varepsilon_{\mu\nu}=-\varepsilon_{\nu\mu}, \
\varepsilon_{01}=+1.
\end{equation}
Alternatively, one may  obtain (\ref{Uti}) by the Lorentz transformation 
of the $2$-velocity $U$. This is achieved by the following replacements imposed 
on (\ref{tprimuu}): $U\to V$ followed by  $(t,x)\to (U^0,U^1)$, and 
$(t',x')\to (U'^0,U'^1)$.
Moving on, we note that (\ref{Uti}) results in 
\begin{align}
U'\cdot U'   =(V\cdot V)(U\cdot U)=
\left\{
\begin{array}{l}
+1  \  \FOR U \AND   V \ \text{subluminal}\\
+1 \  \FOR U \AND V \ \text{superluminal}\\
-1 \  \ \text{otherwise}
\end{array}
\right..
\label{tU2}
\end{align}
Thereby, the relative velocity of  two reference frames is always 
superluminal when, relative to some other reference frame, one of them 
is superluminal while the other one is subluminal.
Otherwise, the discussed relative velocity is subluminal, which 
we find interesting (obvious) when both reference frames are superluminal
(subluminal) with respect to some other reference frame.

Equation (\ref{Uti}) can be cast into the following more familiar form.
With every $2$-vector $U$, we associate the
velocity $u$ via 
\begin{equation}
u=\frac{U^1}{U^0}.
\label{1vel}
\end{equation}
Such a definition follows from the $x'=0$ version of (\ref{tuu}) and it leads to the 
following $U$-parametrization $U=\B{U^0,U^0u}=\pm\B{\gamma(u),\gamma(u)u}$.
Combining (\ref{Uti}) and (\ref{1vel}), we arrive at  
\begin{equation}
u'=\frac{V^0 U^1-V^1 U^0}{V^0U^0-V^1U^1},
\label{util}
\end{equation}
which is antisymmetric with respect to the $U\leftrightarrow V$
transformation
in accordance with standard expectations. We find it curious that 
(\ref{Uti}) does not exhibit such a symmetry.
We note that whenever 
$V\sim(1,v)$ and 
$U\sim(1,u)$, (\ref{util}) leads to
the standard formula (see e.g. \cite{ksiazka,sokolowski})
\begin{equation}
u'=\frac{u-v}{1-u v}.
\label{add}
\end{equation}

Keeping in mind that $V^0=\sgn(V^0)\gamma(v)$, $V^1=\sgn(V^0)\gamma(v)v$, etc.,
we arrive at another representation of (\ref{Uti})
\begin{align}
\label{Utilll}
&V\cdot U\neq0: \ U'=V\cdot V\sgn(V\cdot U)\B{\gamma(u'),\gamma(u')u'},\\
&V\cdot U=0: \ U'=\B{0,\sgn(V^0 U^1)}.
\label{Utilll0}
\end{align}
 Two remarks are in order now.
 
Firstly, we note that 
$V\cdot U=0$ is satisfied by $V=\pm(U^1,U^0)$ or 
equivalently $U=\pm(V^1,V^0)$, which can be easily visualized on spacetime
diagrams 
such as the ones depicted in Fig. \ref{4osie}. 
In the traditional nomenclature,  $V\cdot U=0$ amounts to the well-known 
condition $uv=1$ for $|u'|=\infty$ (see e.g. \cite{ksiazka,sokolowski}),
where $u$ and $v$ are defined as in 
(\ref{1vel}). We would like to stress that the condition $V\cdot U=0$, unlike 
$uv=1$, has clear geometrical meaning in  Minkowski spacetime.
We also note that the right-hand side of (\ref{Utilll0})
implies  $|u'|=\infty$.

 Secondly, (\ref{Utilll}) can be used to argue that 
 reference frames propagating backwards in time 
 inevitably appear in our formalism  when superluminal 
 velocities are considered, which is counterintuitive. 
 Indeed, taking $V=(\gamma(v),\gamma(v)v)$
 and $U=(\gamma(u),\gamma(u)u)$, where $V\cdot U\neq0$ 
 and both $2$-velocities 
 describe reference frames moving forwards in time $t$, we 
 see from (\ref{Utilll}) that $U'$ describes propagation 
 backwards in time $t'$ when\footnote{Strictly speaking, from the perspective of the 
 reference frame $O'$, the reference frame $O''$ propagates backwards 
 in time $t'$ when (\ref{minus1}) holds.}
 \begin{equation}
     \sgn(1-v^2)\sgn(1-uv)=-1,
     \label{minus1}
 \end{equation}
 which  can  be satisfied when at least one 
 of the velocities is superluminal.

Finally, we mention that the transformation 
\begin{equation}
    (t,x) \ \leftrightarrow \ (t',x')
\label{tyghj}
\end{equation}
is induced by 
\begin{equation}
    \B{U^0,U^1} \ \rightarrow \ U\cdot U \B{U^0,-U^1},
    \label{Urec}
\end{equation}
which can be seen as velocity reciprocity in our formalism
(see \cite{Moylan2022} for the recent comprehensive 
discussion of  velocity reciprocity in relativity theories).
For subluminal $U$, (\ref{Urec}) reduces to 
$\B{U^0,U^1} \rightarrow \B{U^0,-U^1}$: the spatial 
component of the $2$-velocity gets flipped.
For superluminal $U$, (\ref{Urec}) reads 
$\B{U^0,U^1} \rightarrow \B{-U^0,U^1}$: the temporal
component of the $2$-velocity gets flipped. 
We mention in passing that  
(\ref{tyghj}) is enforced by the flip of the velocity,
i.e. $u\to-u$, in subluminal and superluminal 
transformations (\ref{superIIIgen}).
This is not the case in our formalism in the superluminal case.

\begin{figure}[t]
\includegraphics[width=\columnwidth]{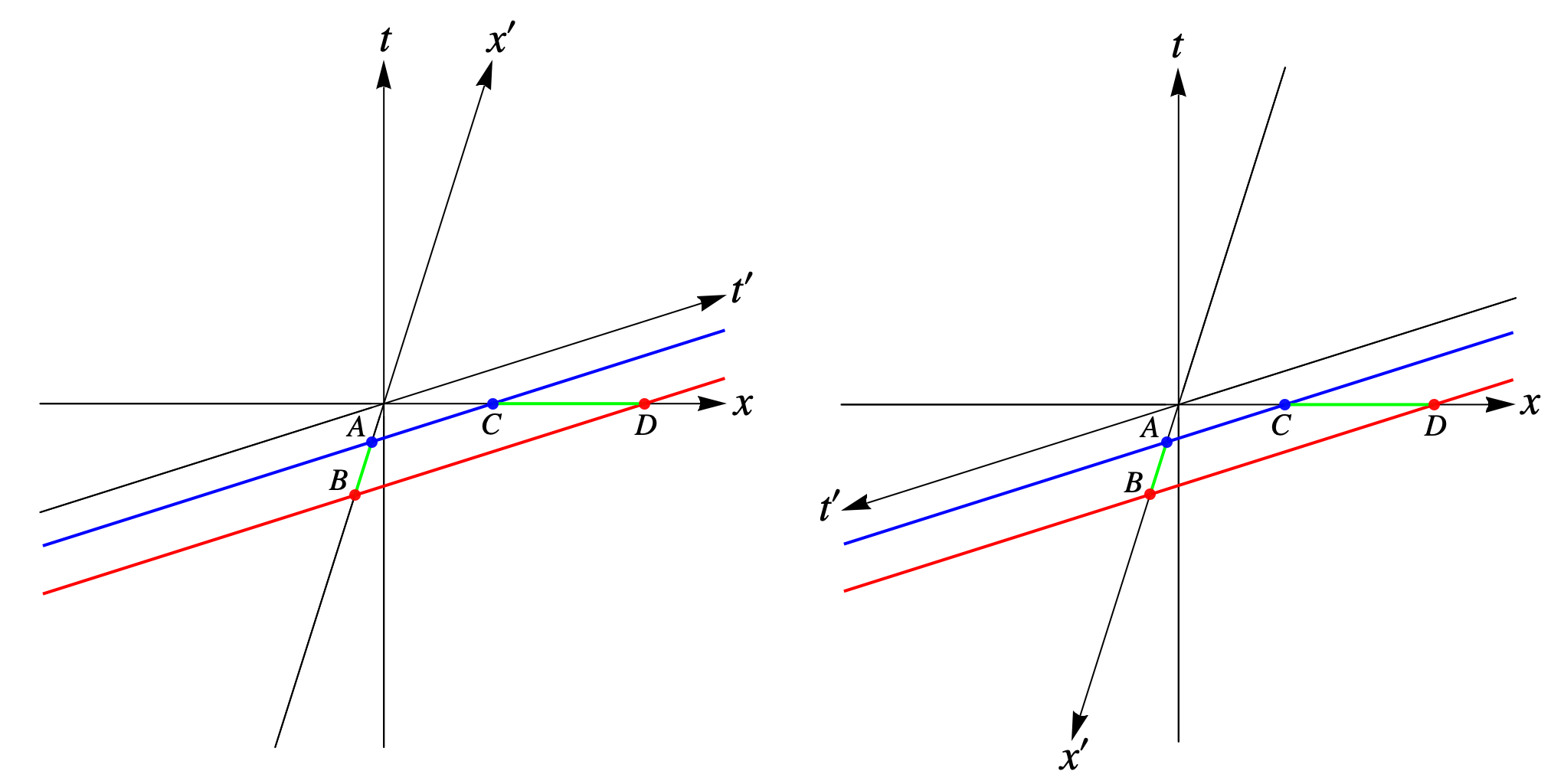}
\caption{The schematic plot illustrating the discussion of length
``contraction'' (see Sec. \ref{length_sec} for the
definition  of the events $A$, $B$, $C$, and $D$). 
The rod is depicted via the thick green line. Its endpoints are marked by blue and red dots
(worldlines of the rod's ends  are shown in the same
colors).
While both panels are 
prepared for the same $U^1/U^0=u>1$, $U^0$ is larger (smaller) than zero in the 
left (right) panel. 
The rod is simultaneously at rest in both 
primed reference frames. Its appearance in the 
unprimed reference frame does not depend on whether the primed reference frame moves
forwards or backwards in time.
}
\label{ustawienia}
\end{figure}
\section{Length ``contraction'' and time ``dilation'' }
\label{length_sec}
We discuss  here further properties of transformation (\ref{tprimuu}) and 
its inverse  (\ref{tuu}), i.e. we again assume that the primed reference frame 
moves with the $2$-velocity $U$ with respect to the unprimed 
reference frame.

{\bf Length ``contraction''.} 
Measuring the length of a rod requires 
that an observer, in his own reference frame, 
simultaneously   determines the  spatial  positions of the
rod's endpoints. 
We consider the rod whose endpoints
are located at $A=(t'_A,x'_A)$ 
and $B=(t'_B,x'_B)$ in the primed 
reference frame and at 
$C=(t_C,x_C)$
and $D=(t_D,x_D)$ in the unprimed
reference frame, respectively. 
We assume that the rod is at rest in the primed reference frame and 
so its proper length $\ell'$ is given by $|x'_A-x'_B|$
even when the events $A$ and $B$ are not simultaneous 
in the primed reference frame.
The  rod's length in the 
unprimed reference frame is $\ell=|x_C-x_D|$ as long as 
$t_C=t_D$.
It follows from (\ref{tprimuu}) that the spatial coordinates
of the events $C$ and $D$, as observed in the primed reference frame,
are 
\begin{equation}
x'_C=U\cdot U (U^0 x_C - U^1 t_C), \ x'_D=U\cdot U (U^0 x_D - U^1 t_D).
\end{equation}
Subtracting these two equations from each other, 
and keeping in mind that 
$x'_C=x'_A$, $x'_D=x'_B$, and $t_C=t_D$, we obtain
\begin{equation}
\ell=\frac{\ell'}{|(U\cdot U)U^0|}=\frac{\ell'}{\gamma(u)}
\end{equation}
via (\ref{Upm}). 
These considerations are illustrated in Fig. \ref{ustawienia}, 
where we set $t'_A=t'_B$  and consider the primed reference frame 
 moving either forwards or backwards in time with a superluminal velocity.

{\bf Time ``dilation''.}
We consider a clock  at rest in the primed reference frame. This clock
measures the time interval $\Delta t'_{AB}=t'_B-t'_A$ between events $A=(t'_A,x'_A)$ and $B=(t'_B,x'_B)$ occurring at 
 $x'_A=x'_B$.
As measured by  the clock resting in the unprimed reference frame, the time interval separating these events is
\begin{equation}
\Delta t_{AB}=t_B-t_A=U^0\Delta t'_{AB}=\sgn(U^0)\gamma(u)\Delta t'_{AB}
\label{DilT}
\end{equation}
according to  (\ref{tuu}) and (\ref{Upm}). 
It is clear from (\ref{DilT}) that the sequence of events $A$ and $B$ is swapped when $\sgn(U^0)=-1$, i.e. when
the primed reference frame is moving backwards in time $t$. Such a conclusion can be easily visualized 
with the help of the diagrams from Fig. \ref{4osie} (one may assume for simplicity
that the events in the primed reference frame take place on the $t'$ axis:  $x'_A=x'_B=0$).

Finally, we note that for  $|u|>\sqrt{2}$:  $\ell>\ell'$ 
(contraction is not always seen) and 
$|\Delta t_{AB}|<|\Delta t'_{AB}|$ (dilation is not always seen).
These remarks explain our use of quotation marks 
around the words contraction and dilation.
We remark that the special role of $u=\sqrt{2}$ in the context of
length ``contraction'' and time ``dilation'' was noted in 
\cite{dragan2008devil}.

\section{Restricted transformations}
\label{5_sec}

We discuss here transformations
restricted to only two (out of four possible) spacetime orientations
of the $2$-velocity $U$. On the one hand, this will give us another 
opportunity to illustrate how our $U$-parametrization works in practice. On the other 
hand, this will allow us to discuss from a different angle
superluminal Lorentz
transformations that were studied in the literature.

We begin the discussion of such transformations from 
(\ref{Itr}), which looks like  a natural extension of the 
standard Lorentz transformation to superluminal velocities. 
In our $U$-parametrization,  (\ref{Itr}) reads
\begin{equation}
u\in\mathbbm{R}\setminus\{1,-1\}: \ U =   \sgn(1-u^2)\B{\gamma(u),\gamma(u)u}.
\label{ItrU}
\end{equation}
Transformations having such a structure, however, 
do not form a group, which can be  shown 
with the help of (\ref{Utilll}). Namely, we choose $u$ and $v$ such that 
$u v\neq1$, which allows us to put
(\ref{ItrU}) and 
$V=\sgn(1-v^2)\B{\gamma(v),\gamma(v)v}$ into (\ref{Utilll}). This leads to
\begin{equation}
U'= \sgn(1-u^2)\sgn(1-uv)\B{\gamma(u'),\gamma(u')u'}=\sgn(1-v^2)\sgn(1-u v)\sgn(1-u'^2)\B{\gamma(u'),\gamma(u')u'},
\label{sdwdewd}
\end{equation}
where the last equality follows from 
\begin{equation}
    \sgn(1-u'^2)=\sgn(1-v^2)\sgn(1-u^2)
\end{equation}
obtained from (\ref{add}).
Result (\ref{sdwdewd}) does not agree with the $u\to u'$ version of (\ref{ItrU})
when (\ref{minus1}) holds.
If both $u$ and $v$ are subluminal, then 
 (\ref{minus1}) cannot be satisfied, which is expected because 
 we deal in such a case with the standard Lorentz transformation.
However, when at least one of these velocities  is superluminal,
then the other can  always be chosen  so as to satisfy 
(\ref{minus1}).
The very same problem  appears
when one uses  the $\gamma(u)\to\sgn(1-u^2)\gamma(u)$ version of (\ref{Itr}),
where 
\begin{equation}
 u\in\mathbbm{R}\setminus\{1,-1\}: \   U =\B{\gamma(u),\gamma(u)u}. 
 \label{ItrUU}
\end{equation}
Thereby, (\ref{Itr}) and its $\gamma(u)\to\sgn(1-u^2)\gamma(u)$ version
break the group property that any Lorentz transformation should satisfy, 
which was  noted in   \cite{Antippa1983}.
In our formalism, the clockwork principle leads to the conclusion 
that superluminal  (\ref{ItrU}) [(\ref{ItrUU})] describes reference frames 
moving only backwards [forwards] in time $t$ and 
either in the positive or negative direction of the $x$ axis 
(this is evident from the fact that 
superluminal $U$ representing (\ref{ItrU}) [(\ref{ItrUU})] are 
depicted in Figs. \ref{4osie}c and \ref{4osie}d
[Figs. \ref{4osie}a and \ref{4osie}b]). 
This observation shows that the restriction to reference frames,
which according to our formalism propagate  
in a fixed direction in time, is insufficient when superluminal 
velocities  are considered.

Then, we take a close look at transformation (\ref{superIIIgen}) employed
in \cite{parker,dragan2008devil,dragan, Antippa1983}.
Transformations having such a structure form a group (see e.g. \cite{dragan} for a recent insight into 
this issue as well as \cite{Antippa1983}). This is seen in our formalism  as follows.
The $U$-parametrization of (\ref{superIIIgen}) is 
\begin{subequations}
\begin{align}
&|u|<1: \ U=\B{\gamma(u),\gamma(u)u},\\
&|u|>1: \ U=\B{\sgn(u)\gamma(u),\gamma(u)|u|},
\label{UIIIb}%
\end{align}
\label{UIII}%
\end{subequations} 
which can be equivalently written as 
\begin{equation}
   u\in\mathbbm{R}\setminus\{1,-1\}: \    U=\sgn(1+u)\B{\gamma(u),\gamma(u)u}
\label{UIV}
\end{equation}
thanks to the compact representation of (\ref{superIIIgen}) proposed in \cite{grudka_2023}.
We choose $u$ and $v$ such that 
$u v\neq1$ and substitute (\ref{UIV}) and $V=\sgn(1+v)\B{\gamma(v),\gamma(v)v}$ into 
(\ref{Utilll}) getting
\begin{equation}
   U'=\sgn(1-v)\sgn(1+u)\sgn(1- uv)\B{\gamma(u'),\gamma(u')u'}=\sgn(1+u')\B{\gamma(u'),\gamma(u')u'},
\label{hjkop}
\end{equation}
where the last equality is obtained from (\ref{add}).
Result (\ref{hjkop}) proves that $U'$ is given by the $u\to u'$ version of 
(\ref{UIV}) when $u v\neq1$. For $uv=1$, one may easily arrive at the 
same conclusion by using  (\ref{Utilll0}) instead of
(\ref{Utilll}).
Moreover, by the same token one may show that 
the $-\sgn(u)\to\sgn(u)$ version of (\ref{superIIIgen}), which is
$U$-parametrized as 
\begin{equation}
    u\in\mathbbm{R}\setminus\{1,-1\}: \  U=\sgn(1-u)\B{\gamma(u),\gamma(u)u}
    =\left\{
    \begin{array}{l}
     |u|<1: \ U=\B{\gamma(u),\gamma(u)u}  \\
    |u|>1: \ U=-\B{\sgn(u)\gamma(u),\gamma(u)|u|}
    \end{array}
    \right.,
    \label{UV}
\end{equation}
also satisfies the 
group property.
Proceeding as above, we note that the clockwork principle leads to 
the observation that superluminal  (\ref{UIV}) [(\ref{UV})] describes reference frames 
moving in the positive [negative] direction of the $x$ axis and 
either forwards or backwards in time $t$:
$U$ corresponding to superluminal (\ref{UIV}) [(\ref{UV})] are shown in 
Figs. \ref{4osie}a and  \ref{4osie}d [Figs. \ref{4osie}b and  \ref{4osie}c].

To fix such a  problematic one-directional movement in space,
one is forced to use the reinterpretation principle,
which states 
that   motion backwards in time $t$ in the positive (negative) direction
of the $x$ axis represents motion forwards in time $t$ in the negative (positive)
direction of the $x$ axis \cite{sudarshan}. Therefore, the reinterpretation principle 
allows one to work with only two orientations of  spacetime axes as in 
Figs. \ref{4osie}a and \ref{4osie}d or 
Figs. \ref{4osie}b and \ref{4osie}c 
(instead of four as in Figs. \ref{4osie}a--\ref{4osie}d).
However, one should be aware of the fact that by choosing (\ref{superIIIgen})
as in \cite{parker,dragan2008devil,dragan, Antippa1983}, 
one deals with the situation, where the superluminal observer moving 
in the positive (negative) direction of the $x$ axis uses a clock with a normal (inverted) mechanism
(the hands of the clock are moving clockwise (counterclockwise) in these two cases,
so to speak).  
Similarly, by using (\ref{superIIIgen}) subjected to the
$-\sgn(u)\to\sgn(u)$ replacement, 
one assumes that
the superluminal observer uses a clock with a normal (inverted) mechanism
when it moves in the negative  (positive) direction 
of the $x$ axis.
We  see such dependence of the clock's
mechanism on the direction of motion as a counterintuitive  feature.

\section{Summary}
\label{6_sec}
In the spirit of the theory of relativity, we have considered reference frames moving in all possible directions
in space and time. In particular, this implies that we have taken into account   reference frames 
moving both forwards and backwards in time. While the exploration of both options
does not seem to be necessary 
in the subluminal context, the situation is far less clear in the superluminal context, where 
one is bound to encounter counterintuitive features and 
lack of experimental data leaves  various  possibilities open.
We hope that our work will stimulate the discussion of this issue.

In our formalism, we have parameterized Lorentz transformations via the $2$-velocity of a reference frame and
introduced the clockwork principle to give physical meaning to this 
quantity. We have 
studied then the group property of such transformations and re-examined basic effects such as 
the length ``contraction'' and the time ``dilation''. The new formalism has been then compared to the 
standard approach  discussed in  \cite{parker,dragan2008devil,dragan, Antippa1983}. In the course
of these studies,
we have identified counterintuitive features both in our formalism and 
in the above-mentioned standard approach (see  the discussion around
(\ref{minus1}) and by the end of Sec. \ref{5_sec}).
In fact, we believe that 
any approach to superluminal systems is 
going to encounter some conceptual difficulties needing detailed discussion.
Such a remark partly motivates
research pursuits discussed in this work, which we hope 
gives a non-standard perspective on superluminal physics.
Finally, we note that it is of interest to extend the formalism presented in 
this work to higher dimensional spacetimes, particularly in light of recent developments 
presented in  \cite{Dragan1plus3}.


%

\end{document}